\documentclass[11pt,a4paper,english,superscriptaddress,aps,nofootinbib]{revtex4}
\usepackage{amsmath,amssymb,graphicx}
\makeatletter
\usepackage{babel}
\usepackage[active]{srcltx}
\usepackage{graphicx,color}
\usepackage{changebar}
\usepackage{hyperref}
\hypersetup{
	colorlinks=true,
	urlcolor= blue,
	citecolor=red,
	linkcolor= blue,
	bookmarks=true,
	bookmarksopen=false,
}

\usepackage[dvipsnames]{xcolor}
\usepackage[T1]{fontenc}
\usepackage{ae}
\usepackage[ansinew]{inputenc}
\usepackage{amsmath}
\usepackage{braket}
\usepackage{mathtools}
\usepackage{slashed}
\usepackage{empheq}
\usepackage{tikz}
\usetikzlibrary{decorations.pathmorphing}
\usepackage{graphicx}
\usepackage{amsfonts}
\usepackage{amsmath}
\newcommand{\be}{\begin{equation}}
\newcommand{\ee}{\end{equation}}
\newcommand{\bea}{\begin{eqnarray}}
\newcommand{\eea}{\end{eqnarray}}

\begin{document}
\title{Nonlinear $\sigma$-models in the Eddington-inspired Born-Infeld Gravity}
\author{J. R. Nascimento}
\email[]{jroberto@fisica.ufpb.br}
\affiliation{Departamento de F\'{\i}sica, Universidade Federal da 
	Para\'{\i}ba,\\
	Caixa Postal 5008, 58051-970, Jo\~ao Pessoa, Para\'{\i}ba, Brazil}

\author{Gonzalo J. Olmo}
\email[]{gonzalo.olmo@uv.es}
\affiliation{Departament de F\'{i}sica Te\`{o}rica and IFIC, Centro Mixto Universitat de
	Val\`{e}ncia - CSIC,\\
	Universitat de Val\`{e}ncia, Burjassot-46100, Val\`{e}ncia, Spain}

\author{P. J. Porf\'{i}rio}\email[]{fepa@sas.upenn.edu}
\affiliation{Departament of Physics and Astronomy, University of Pennsylvania, Philadelphia, PA 19104, USA}
\affiliation{Departamento de F\'{\i}sica, Universidade Federal da 
	Para\'{\i}ba,\\
	Caixa Postal 5008, 58051-970, Jo\~ao Pessoa, Para\'{\i}ba, Brazil}

\author{A. Yu. Petrov}
\email[]{petrov@fisica.ufpb.br}
\affiliation{Departamento de F\'{\i}sica, Universidade Federal da 
	Para\'{\i}ba,\\
	Caixa Postal 5008, 58051-970, Jo\~ao Pessoa, Para\'{\i}ba, Brazil}

\author{A. R. Soares}
\email[]{adriano2da@gmail.com}
\affiliation{Departamento de F\'{\i}sica, Universidade Federal da 
	Para\'{\i}ba,\\
	Caixa Postal 5008, 58051-970, Jo\~ao Pessoa, Para\'{\i}ba, Brazil}

\begin{abstract}
In this paper we consider two different nonlinear $\sigma$-models minimally coupled to Eddington-inspired Born-Infeld gravity. We show that the resultant geometries represent minimal modifications with respect to those found in GR, though with important physical consequences. In particular, wormhole structures always arise, though this does not guarantee by itself the geodesic completeness of those space-times. In one of the models, quadratic in the canonical kinetic term, we identify a subset of solutions which are regular everywhere and are geodesically complete. We discuss characteristic features of these solutions and their dependence on the relationship between mass and global charge.

	\end{abstract}

\maketitle
% !TeX spellcheck = en_GB 
 \section{INTRODUCTION}
It is well established that  topological defects could arise in the primordial universe due to phase transitions occurring during rapid expansion and cooling periods \cite{Kibble,GM}. The most studied types of defects are domain walls, cosmic strings and global monopoles (GM) characterized by spontaneous breaking {of the symmetries} $\mathbb{Z}_2$, $SO(2)$, and $SO(3)$ \cite{Kibble}. The last ones are of special interest  because they do not require the introduction of gauge fields, hence allowing their exploration via gravitational interactions. In this context, Barriola and Vilenkin obtained the first solution describing the space-time geometry outside a GM core \cite{GM}. Assuming very light GMs, one finds that their Newtonian potential tends to zero at astrophysical scales, yielding negligible gravitational attraction. Nonetheless, the space-time still presents a solid deficit angle able to produce a certain deflection of light rays. In \cite{GM}, the authors also called attention to the case where the Schwarzschild radius  of the defect is much larger than the radius $\delta$ of the core ($M\gg\delta$), which can be interpreted  as a black hole with GM charge \cite{Pramana}. As a consequence of the presence of this charge, both the event horizon and light deflection -- in the weak and strong field regime -- increase in comparison with the Schwarzschild case \cite{Man2011}. However, just like the majority of other classic black hole solutions of General Relativity (GR), the solution is singular at the origin.

Despite the success of GR in weak and also strong field astrophysical scenarios, the theory suffers from a severe conceptual limitation due to the possibility of engendering singularities, i.e., regions with incomplete geodesics. For example, in  the case of the Schwarzschild solution, any geodesic (or arbitrary null or timelike trajectory) that crosses the event horizon is inevitably headed towards $r=0$, and ends there, with no possible extension beyond. In practical terms, this means that information and observers simply vanish when $r=0$ is reached, leading to an absurd situation in which nothing is observable by anybody. When a theory yields absurd answers  to physical questions (such as infinite values for physical magnitudes or the impossibility of performing measurements) it is evident that the questions being posed lie beyond that theory's capabilities, and an improved description is necessary. From this viewpoint, we should revisit the results predicted by GR in such scenarios from the perspective of an alternative theory of gravity. Alternatives to GR have also become very popular recently for very different reasons related to cosmological observations \cite{cosmology}.

 In this work we are interested in the so-called Eddington-inspired Born-Infeld modification of gravity (EiBI gravity for short) \cite{Vollick:2003qp,banados2010eddington}. The structure of the gravitational Lagrangian in this theory is inspired by the non-linear electrodynamics of Born and Infeld \cite{born1934foundations, eddington1920mathematical}, and is formulated in a metric-affine approach to avoid the higher-derivative equations that typically appear in the standard metric formulation.  The EiBI theory provides, already at the classical level,  interesting solutions like regular black holes, wormholes, nonsingular cosmologies, and many other results without requiring exotic matter sources (for a review on EiBI and its applications see \cite{BeltranJimenez:2017doy}). In the absence of matter, EiBI is equivalent to GR plus possibly an effective cosmological constant. However it differs from GR in the presence of matter. In this sense, in the innermost regions of compact objects, where the energy density reaches its highest values, new effects arise that can avoid geodesic incompleteness in some cases \cite{Olmo:2015dba,Bazeia:2015uia,Olmo:2015bya,Bambi:2015zch}. Here we will explore how EiBI combined with global monopoles modify the internal structure of black holes.

Different studies of the space-time generated by GMs have already been carried out in the context of alternative theories of gravity. In \cite{barros1997global}, Barros and Romero investigated GMs in the weak field approximation of Brans-Dicke theories. Later, Carames \textit{et al.} \cite{carames2011gravitational} considered GMs in the context of  $f(R)$ gravity, which motivated many other studies \cite{carames2012motion, morais2012gravitational, man2013thermodynamic, man2015analytical, carames2017f}. However, at least at the classical level, these works failed to get rid of the singularity. In our previous paper \cite{Palatinef(R)mono}, the GM in a metric-affine $f(R)$ theory was studied and it was shown that the model supports both singular and regular (geodesically complete) solutions.  In \cite{lambaga2018gravitational}, Lambaga and Ramadhan investigated black holes with GM topological charge in EiBI, also in the metric-affine formalism. They explored only the external region of the solution, favoring a positive value for the theory's parameter $\epsilon$ which, in this case, leads to a singular solution. In this paper we will consider also the $\epsilon<0$ case having as matter source different types of nonlinear $\sigma$-models. In this context, the GM spacetime will be an exact solution of the field equations \cite{Hongwei}. We will see that for negative values of the EiBI parameter we can have geodesically complete solutions with distinct characteristics depending on the  GM charge-to-mass ratio. Thus, we show that unlike in GR, the presence of GM charge in the EiBI gravity dramatically changes the solution profile in the innermost regions, opening up new possibilities to improve the classical behavior of GR. 

The paper is organized as follows: In section \ref{EiBI}, we review the field equations motivated by the more general approach systematized in \cite{olmo2011}. This discussion is appropriate because of the energy-momentum tensor structure we are going to use.  In Section \ref{SEiBI}, we obtain the form of the metric correspondent to a specific kind of anisotropic fluid. In section \ref{sigmamodel}, we introduce the global monopole as a matter source of NLSM whose energy-momentum tensor fits in the structure considered in section \ref{SEiBI}. There, we consider particular models, namely, canonical and power law fields, and then we examine the corresponding solutions. 
%Next, in section \ref{correspondencia}, we lay out a correspondence between NLSM and non-linear electrodynamics in EiBI gravity. 
Finally, we present our conclusions in section \ref{conclusao}. In the Appendix, we give a general description of the topologically charged Ellis wormhole.

\section{Born-Infield Gravity}\label{EiBI}

The action of the EiBI theory can be written as
\begin{equation}\label{acao-BI}
S_{EiBI}=\frac{1}{8\pi G\epsilon}\int d^4x\left[ \sqrt{-|g_{\mu\nu}+\epsilon R_{(\mu\nu)}(\Gamma)|}-\lambda\sqrt{-|g_{\mu\nu}|}\right]+S_m[g_{\mu\nu}, \Phi].
\end{equation} 
Here $G$ is Newton's gravitational constant, $\epsilon$ is a parameter  with dimension of area that controls the nonlinearity of the theory, vertical bars denote matrix determinant, the Ricci tensor $R_{(\mu\nu)}(\Gamma)$ is symmetrized to avoid  ghost-like degrees of freedom \cite{BeltranJimenez:2019acz} and is constructed assuming that the connection $\Gamma$ is {\it a priori} independent of the metric $g_{\mu\nu}$. The term $S_{m}[g_{\mu\nu},\Phi]$ represents the action of the matter fields  $\Phi$. In general, the constant $\lambda$ defines an effective cosmological constant $\lambda=1+\epsilon\Lambda $ but here it will be set to $\lambda=1$ for simplicity. In terms of the definitions
\begin{equation}\label{hmn}
	h_{\mu\nu}=g_{\mu\nu}+\epsilon R_{\mu\nu},
\end{equation}
and $\kappa^2=8\pi G$, the action (\ref{acao-BI}) reads as
\begin{equation}
	S_{BI}=\frac{1}{\epsilon\kappa^2}\int\left[\sqrt{-h}-\sqrt{-g}\right]d^4x+S_m[g_{\mu\nu},\Phi],
\end{equation}  	
where $h$ and $g$ are, respectively, the determinants of the symmetric tensors $h_{\mu\nu}$ and $g_{\mu\nu}$. The variation of the action with respect to the metric $g_{\mu\nu}$ and the connection $\Gamma^\alpha_{\mu\nu}$ leads to field equations 
\begin{equation}\label{EoM1}
	\sqrt{-h}h^{\mu\nu}=\sqrt{-g}\left(g^{\mu\nu}-\epsilon\kappa^2T^{\mu\nu}\right),
\end{equation}
\begin{equation}\label{EoM2}
	\nabla_{\mu}\left(\sqrt{-h}h^{\alpha\beta}\right)=0,
\end{equation}
where the energy-momentum tensor is given by $T^{\mu\nu}=\frac{2}{\sqrt{-g}}\frac{\delta S_m}{\delta g_{\mu\nu}}$, and $h^{\alpha\beta}$ formally denotes the inverse of $h_{\mu\nu}$. The form of (\ref{EoM2}) assumes vanishing torsion, though for minimally coupled bosonic matter fields this fact is irrelevant (see \cite{Afonso:2017bxr} for details). 
From (\ref{EoM2}), we conclude that the connection is simply the Levi-Civita connection of the auxiliary metric $h_{\mu\nu}$, namely, $\Gamma_{\mu\nu}^{\alpha}=\frac{1}{2}h^{\alpha\lambda}\left(\partial_\mu h_{\lambda\nu} +\partial_\nu h_{\lambda\mu}-\partial_\lambda h_{\mu\nu}\right)$. The explicit form of $h_{\mu\nu}$ can be obtained through the relation $\epsilon R_{\mu\nu}(h)=h_{\mu\nu}-g_{\mu\nu}$ that follows from Eq. (\ref{hmn}) once $g_{\mu\nu}$ is written in terms of $h_{\mu\nu}$ and the stress-energy tensor of the matter fields. For this purpose, we propose that the auxiliary metric $h_{\mu\nu}$ and the physical metric $g_{\mu\nu}$ be related by means of a deformation matrix $\Omega^{\alpha}_{\phantom{\alpha}\nu}$ according to
\begin{equation}\label{h-conformal-g}
	h_{\mu\nu}=g_{\mu\alpha}\Omega^{\alpha}_{\phantom{\alpha}\nu} \ , \ \quad h^{\mu\nu}=\left(\Omega^{-1}\right)^{\mu}_{\phantom{\mu}\alpha}g^{\alpha\nu},
\end{equation}
{and} use these relations in Eq. (\ref{EoM1}) to obtain
\begin{equation}\label{Matrix-deformation}
	\sqrt{|\Omega|}\left(\Omega^{-1}\right)^{\mu}_{\phantom{\mu}\nu}=\delta^{\mu}_{\phantom{\mu}\nu}-\epsilon\kappa^2T^{\mu}_{\phantom{\mu}\nu} \ , 
\end{equation}
which shows that the deformation that relates the metrics is determined by the local stress-energy densities. 

Now let us return to Eq. (\ref{hmn}). Contracting  the equation $\epsilon R_{\alpha\nu}(h)=h_{\alpha\nu}-g_{\alpha\nu}$ with $h^{\alpha\mu}$, we find that $\epsilon R^{\mu}_{\phantom{\mu}\nu}(h)=\delta^{\mu}_{\phantom{\mu}\nu}-(\Omega^{-1})^{\mu}_{\phantom{\mu}\nu}$ and then, from Eq. (\ref{Matrix-deformation}), we {arrive at}
\begin{equation}\label{E-EoM}
	R^{\mu}_{\phantom{\mu}\nu}(h)=\frac{\kappa^2}{\sqrt{|\Omega|}}\left[\frac{(\sqrt{|\Omega|}-1)}{\epsilon\kappa^2}\delta^{\mu}_{\phantom{\mu}\nu}+T^{\mu}_{\phantom{\mu}\nu}\right].
\end{equation}
From this set of partial differential equations one can obtain $h_{\mu\nu}$ and, then, through (\ref{h-conformal-g}) we can find the physical metric $g_{\mu\nu}$. From equation (\ref{Matrix-deformation}) one can conclude that the vacuum solutions in this model are the same as in the GR. Therefore, the presence of matter fields will be necessary to achieve novelties in the gravitational dynamics, specially in the innermost regions of black holes.

\section{ Static spherically symmetric solutions for anisotropic fluids}\label{SEiBI}
Given that the stress-energy tensor determines the deformation matrix $\Omega^{\alpha}_{\phantom{\alpha}\nu}$, it is natural to assume that this matrix has the same algebraic structure as $T^{\alpha}_{\phantom{\alpha}\nu}$. For this reason, if one considers a generic matter source with the structure  of an anisotropic fluid of the form
 \begin{equation}\label{ten}
T^{\mu}_{\phantom{\mu}\nu}=\text{diag} (-\rho,-\rho, P_{\theta},P_{\theta}).
 \end{equation}
it follows that 
\begin{equation}\label{Omegaline}
	\Omega^{\mu}_{\phantom{\mu}\nu}=\text{diag}\left[\Omega_{+}(r), \Omega_{+}(r),\Omega_{-}(r),\Omega_{-}(r)\right],
\end{equation}
where $\Omega_{\pm}$ are functions determined by Eq. (\ref{Matrix-deformation}) which have the explicit form
\begin{equation}\label{omegas}
\Omega_{-}=1+\epsilon\kappa^2 \rho,\qquad\Omega_{+}=1-\epsilon\kappa^2 P_{\theta}.
\end{equation}
As we can clearly see, if there is no matter or $\epsilon\to 0$, the deformation matrix becomes the identity and, therefore, $h_{\mu\nu}$ and $g_{\mu\nu}$ become identical. In regions with nonvanishing energy density, however, they will be different. 
For static and spherically symmetric solutions, we can adopt the following ansatz for the line element characterized by $h_{\mu\nu}$:
 \begin{equation}
 d\tilde{s}^2\equiv h_{\mu\nu}dx^\mu dx^\nu=-A(x)e^{2\Phi(x)}dt^2+\frac{dx^2}{A(x)}+\tilde{r}^2(x)(d\theta^2+\sin^2\theta d\phi^2).
 \end{equation}
 Calculating $R^{\mu}_{\phantom{\mu}\nu}(h)$, we arrive at
 \begin{equation}\label{Rtx}
 R^{t}_{\phantom{t}t}(h)-R^{x}_{\phantom{x}x}(h)=\frac{2}{\tilde{r}}\left(\frac{d^2\tilde{r}}{dx^2}-\frac{d\Phi}{dx}\frac{d\tilde{r}}{dx}\right),
 \end{equation}
 \begin{equation}
 R^{\theta}_{\phantom{\theta}\theta}(h)=\frac{1}{\tilde{r}^2}\left[1-\tilde{r}\frac{d\tilde{r}}{dx}\left(A\frac{d\Phi}{dx}+\frac{dA}{dx}\right)-A\left(\tilde{r}\frac{d^2\tilde{r}}{dx^2}+\left(\frac{d\tilde{r}}{dx}\right)^2\right)\right].
 \end{equation}
 Since $T^{t}_{\phantom{t}t}=T^{x}_{\phantom{x}x}$, it follows from Eq. (\ref{Rtx}) that $\left(\frac{d^2\tilde{r}}{dx^2}-\frac{d\Phi}{dx}\frac{d\tilde{r}}{dx}\right)=0$. Without loss of generality, this result allows us to take $\Phi(x)=0$ and $\tilde{r}=x$. So we can write the line element for $h_{\mu\nu}$ as
 \begin{equation}\label{metricah}
 d\tilde{s}^2=-A(x)dt^2+\frac{dx^2}{A(x)}+x^2(d\theta^2+\sin^2\theta d\phi^2).
 \end{equation}
 From  Eqs. (\ref{h-conformal-g}) and (\ref{Omegaline}), one can then write the line element for $g_{\mu\nu}$ as
 \begin{equation}\label{metrica}
 ds^2=-\frac{A(x)}{\Omega_{+}}dt^2+\frac{1}{\Omega_{+}A(x)}dx^2+r^2(x)(d\theta^2+\sin^2\theta d\phi),
 \end{equation}
 where $r^2(x)=\frac{x^2}{\Omega_{-}}$. Therefore, Eq. (\ref{omegas}) implies
\begin{equation}\label{rel-x-r}
	x^2=r^2+\epsilon\kappa^2r^2\rho.
\end{equation}
With the simplifications made above, the component $R^{\theta}_{\phantom{\theta}\theta}(h)$  now becomes
 \begin{equation}\label{teta}
 R^{\theta}_{\phantom{\theta}\theta}(h)=\frac{1}{x^2}\left(1-A-x\frac{dA}{dx}\right).
 \end{equation}
Let us now choose the ansatz
 \begin{equation}\label{ansatzparaA}
 A(x)=1-\frac{2M(x)}{x},
 \end{equation}
 from which we are left with
\begin{equation}
 R^{\theta}_{\phantom{\theta}\theta}(h)=\frac{2}{x^2}\frac{dM}{dx}.
\end{equation}
 Now we can express the l.h.s. of the {above} equation through Eq. (\ref{E-EoM}), and we find
\begin{equation}\label{dM}
	\frac{2}{x^2}\frac{dM}{dx}=\frac{\kappa^2}{\sqrt{|\Omega|}}\left(\frac{\sqrt{|\Omega|}-1}{\epsilon\kappa^2}+P_{\theta}\right),
\end{equation}
which eventually leads to 
\begin{equation}
	\frac{dM(x)}{dx}=\frac{\kappa^2r^2\rho}{2}.
\end{equation}
So, we have
\begin{equation}\label{A}
	A(x)=1-\frac{2M_0}{x}-\frac{\kappa^2}{x}\int r^2\rho dx,
\end{equation}
 where $M_0$ is an integration constant. Now, using Eq. (\ref{rel-x-r}), the general solution {can be expressed} in terms of either $x$ or $r$, a choice that should be made upon  convenience. In the next section, we will present a class of nonlinear $\sigma$-models whose corresponding stress-energy tensor fits into the algebraic structure $T^{\mu}_{\phantom{\mu}\nu}=\text{diag} (-\rho,-\rho, P_{\theta},P_{\theta})$. Hence, from Eqs. (\ref{rel-x-r}) and (\ref{A}) we can get the space-time metric corresponding to the particular matter source considered.

\section{Nonlinear $\sigma$-model}\label{sigmamodel}

The action of a generic K-monopole \cite{Babichev2006} is given by
\begin{equation}\label{acao1}
	S=\int\left[\mathcal{K}(X)-\frac{\lambda}{4}(\vec{\Phi}\cdot\vec{\Phi}-\eta^2)^2\right]\sqrt{-g}d^4x,
\end{equation}
 where $\mathcal{K}(X)$ is a functional of the canonical term $X=\frac{1}{2}\partial_{\mu}\vec{\Phi}\cdot\partial^{\mu}\vec{\Phi}$, and $\vec{\Phi}\equiv\{\Phi^i\}$ corresponds to a triplet of coupled real scalar fields. The model (\ref{acao1}) displays spontaneous symmetry breaking $O(3)\rightarrow U(1)$. The constants $\lambda$ and $\eta$ are, respectively, the coupling constant and the energy scale of the spontaneous symmetry breaking. The specific functional form $\mathcal{K}(X)$ is chosen in such a way that asymptotically the canonical term prevails, thus avoiding the so-called ``zero-kinetic problem'' \cite{Babichev2006,Jin2007}:
\begin{equation}
\mathcal{K}(X)=\left\{\begin{array}{rc}
	-X,&\mbox{if}\quad X\ll 0,\\
	X^{\alpha}, &\mbox{if}\quad X\gg0.
	\end{array}\right\}, 
\end{equation}	
where $\alpha$ is some constant. In the NLSM, the scalar field $\vec{\Phi}$ {should satisfy the  restriction given} by the following equation
	\begin{equation}\label{va}
	 \vec{\Phi}\cdot\vec{\Phi}=\eta^2,
	 \end{equation}
which defines a manifold, in this case a $2-$sphere ($S^2$), in the internal parameters space (moduli space). Such a manifold is known as vacuum manifold, and a particular choice of the parameters on this manifold leads to a spontaneous symmetry breaking.
Note, in this sense, that the constant $\lambda$ in the action (\ref{acao1}) is a Lagrange multiplier. In the vacuum manifold we can define a set of local coordinates $\lbrace\phi^{a}\rbrace$, where $a$ runs from $1$ to $2$, so that $\vec{\Phi}=\vec{\Phi}(\phi^a)$. In order to shed light on this, we can proceed as in \cite{Prasetyo2017} and rewrite the action as  follows:
\begin{equation}
S=\int\mathcal{K}(X)\sqrt{-g}d^4x,
\end{equation}
  where $X=\frac{1}{2}\eta^2\xi_{ij}\partial_{\mu}\phi^i\partial^{\mu}\phi^j$. The quantity $\xi_{ij}$ is the metric of the 2-dimensional Riemannian vacuum manifold:  $\xi_{ij}=\frac{\partial\vec{\Phi}}{\partial\phi^i}\frac{\partial\vec{\Phi}}{\partial\phi^j}$. The field equation and the energy-momentum tensor $T_{\mu\nu}=-\frac{2}{\sqrt{-g}}\frac{\delta S}{\delta g^{\mu\nu}}$ are, respectively, 
 \begin{equation}\label{EoM}
 	\frac{1}{\sqrt{-g}}\partial_\mu\left[\sqrt{-g}\eta^2\mathcal{K}_X\xi_{bi}\partial^{\mu}\phi^i\right]-\frac{\mathcal{K}_X}{2}\eta^2\partial_{\mu}\phi^i\partial^{\mu}\phi^j\frac{\partial\xi_{ij}}{\partial\phi^b}=0;
 \end{equation}
 \begin{equation}
T^{\mu}_{\phantom{\mu}\nu}=\delta^{\mu}_{\nu}\mathcal{K}-\eta^2\mathcal{K}_{X}\partial^{\mu}\phi^a\partial_{\nu}\phi^a.
 \end{equation}
 The index $X$ in $\mathcal{K}_X$ denotes a derivative with respect to $X$. For a spherically symmetric metric, that is, $ds^2=-A(r)dt^2+B(r)dr^2+C(r)(d\theta^2+\sin^2\theta d\varphi^2)$, the ansatz $\phi^1=\theta$ and $\phi^2=\varphi$ identically satisfies the field equations (\ref{EoM}) when
 \begin{equation}
\xi_{ij}=\begin{bmatrix}
1&0\\0&\sin^2\theta
\end{bmatrix}
 \end{equation}
 \cite{Prasetyo2017, Gell}. This ansatz implies $T^0_{\phantom{0}0}$=$T^1_{\phantom{1}1}=\mathcal{K}(X)$, and $T^2_{\phantom{2}2}$=$T^3_{\phantom{3}3}=\mathcal{K}(X)-X\mathcal{K}_X$. Therefore,
 \begin{equation}\label{get}
 T^{\mu}_{\phantom{\mu}\nu}=\text{diag}\left(\mathcal{K}, \mathcal{K}, \mathcal{K}-X\mathcal{K}_{X}, \mathcal{K}-X\mathcal{K}_{X}\right).
 \end{equation}
 Comparing Eq. (\ref{get}) with Eq. (\ref{ten}), we then have
 \begin{equation}\label{energiamomento}
 	\rho=-\mathcal{K}\quad\text{and}\quad P_{\theta}=\mathcal{K}-X\mathcal{K}_{X},
 \end{equation}
 where $X=\frac{\eta^2}{r^2}$. Since the functional form $\mathcal{K}(X)$ is typically a given function, it is possible to get the metric generated by this matter source through { Eqs.} (\ref{rel-x-r}) and (\ref{A}). In the next sections, we consider {several} known forms for $\mathcal{K}(X)$ whose studies have already been performed in the context of GR. We start by considering the canonical case and then go to more complex ones.

 \subsection{Canonical case: $\mathcal{K}=-X$}

According to (\ref{energiamomento}), for this model we have $\rho=\frac{\eta^2}{r^2}$ and $P_{\theta}=0$, which turns (\ref{rel-x-r}) into
 \begin{equation}\label{r-x}
 r^2=x^2-\epsilon\kappa^2\eta^2.
 \end{equation}
 From this it is easy to see that if $\epsilon<0$, then $r^2(x)$ attains a minimum $r^2_{min}=\epsilon\kappa^2\eta^2$ at $x=0$, thus signaling the presence of a wormhole. If $\epsilon>0$ it is $x^2(r)$ which has a minimum, pointing towards the existence of a wormhole in the auxiliary geometry associated to $h_{\mu\nu}$.  Substituting $\rho$ in (\ref{A}), we have $A=1-\kappa^2\eta^2-\frac{2M_0}{x}$, so the general solution is given by
 \begin{equation}\label{metrica-x}
 ds^2=-\left(1-\kappa^2\eta^2-\frac{2M_0}{x}\right)dt^2+\left(1-\kappa^2\eta^2-\frac{2M_0}{x}\right)^{-1}dx^2+(x^2-\epsilon\kappa^2\eta^2)(d\theta^2+\sin^2\theta d\phi^2), 
 \end{equation}
 which in terms of $r$ becomes
 \begin{eqnarray}\label{Ac}
 	ds^2&=&-\left(1-\kappa^2\eta^2-\frac{2M_0}{\sqrt{r^2+\epsilon\kappa^2\eta^2}}\right)dt^2+\frac{r^2}{r^2+\kappa^2\epsilon\eta^2}\left(1-\kappa^2\eta^2-\frac{2M_0}{\sqrt{r^2+\epsilon\kappa^2\eta^2}}\right)^{-1}dr^2+\nonumber\\ &+&
	r^2(d\theta^2+\sin^2\theta d\phi^2).
 \end{eqnarray}
Despite the nonlinear dynamics of the EiBI theory, the simplicity of this solution is remarkable, since it is identical to that found in GR up to the constant shift $-\epsilon\kappa^2\eta^2$ characterizing an angular deficit.
 It was first obtained by Lambaga and Ramadhan \cite{lambaga2018gravitational}, who focused on the $\epsilon>0$ case, though the internal  wormhole geometry was overlooked. It is worth noting that for $M_0=0$ the solution describes an Ellis-like wormhole with topological charge (see the Appendix). On the other hand, focusing on $\epsilon<0$, if $M_0\neq0$,  one finds that the solution is geodesically incomplete. {Indeed,} considering the equatorial plane $\theta=\frac{\pi}{2}$, radial geodesics satisfy the following equation \cite{olmo2011}: 
 \begin{equation}\label{Eq}
\left(\frac{dx}{d\lambda}\right)^2=\Omega_{+}^2E^2-A(x)\Omega_{+}\left(\frac{L^2}{r^2(x)}+k\right) \ ,
 \end{equation}
where $E$, $L$ and $\lambda$ are, respectively, energy, angular momentum, and affine parameter associated with the geodesic, with  $k=+1, 0, -1$ standing for the time-like, null, and space-like geodesics, respectively. To illustrate this, let us consider a time-like geodesic. From (\ref{metrica-x}), near the origin we have
 \begin{equation}
 	\left(\frac{dx}{d\lambda}\right)^2\approx E^2-V_{eff} \ ,
 \end{equation}  
 where  $V_{eff}=-\frac{2M_0}{x}\left(\frac{L^2}{|\epsilon| \kappa^2\eta^2}+k\right)$ plays the role of an effective potential for a particle of energy $E^2$. For a particle on the $x>0$ region,  this effective potential is negative and represents a divergent attractive force as it approaches the center.  However, as soon as it crosses the $x=0$ boundary, the particle finds an infinite potential barrier which makes the right-hand side negative and is inconsistent with the positivity of the left-hand side. Thus, particles are accelerated towards $x\to 0^+$ but are suddenly smashed at $x=0$, with no possibility of going through to the other side. Those geodesics, therefore, are incomplete.  Had we considered initially particles coming from $x<0$ towards $x\to 0^-$, then such particles would approach the center up to a minimal distance, at which $E^2=V_{eff}$, bouncing back safely to the $x<0$ region again.  Null radial geodesics, on the contrary, do not have any problem in going through the wormhole from either side. The geometry, however, must be regarded as singular. 
  
\subsection{Power law case: $\mathcal{K}(X)=-X-\beta X^2$}\label{leidepotencia}
Let us now consider the power law model
\begin{equation}\label{modelo}
	\mathcal{K}(X)=-X-\beta X^2, 
\end{equation}
with $\beta>0$.  From the  GR perspective this model was considered in \cite{Jin2007} and  \cite{Prasetyo2017}, the latter being in higher dimensions. Note that in the limit $\beta\to0$ we get the canonical model. So from (\ref{modelo}) and (\ref{energiamomento}), we have 
\begin{equation}
	\rho=\frac{\eta^2}{r^2}+\beta\frac{\eta^4}{r^4}\quad\text{and}\quad P_{\theta}=\beta\frac{\eta^4}{r^4}.
	\label{matter}
\end{equation}
If we define $Q^2\equiv \beta\eta^4$ we will have $\rho=\frac{\eta^2}{r^2}+\frac{Q^2}{r^4}$, which represents the sum of the energy densities produced by a GM (outside of its core)  plus an electric charge $Q$ \cite{Nordstrom-olmo}. In this way we can interpret the NLSM (\ref{modelo}) as engendering an electric charge $Q$ plus a GM charge $\eta$. Individually, each of these cases has been studied in the context of EiBI gravity \cite{olmo2011, lambaga2018gravitational}. 

Using (\ref{rel-x-r}) it is easy to find the dependence of $r(x)$ with $x$, which becomes
\begin{equation}\label{cor}
	r^2=\frac{(x^2-\epsilon\kappa^2\eta^2)}{2}+\frac{1}{2}\sqrt{(x^2-\epsilon\kappa^2\eta^2)^2-4\epsilon \kappa^2Q^2} \ .
\end{equation}
Inserting now the density function $\rho=\frac{\eta^2}{r^2}+\frac{Q^2}{r^4}$ in Eq. (\ref{A}), one finds
\begin{equation}\label{eq:A(x)}
A(x)=1-\kappa^2\eta^2-\frac{2M_0}{x}-\frac{\kappa^2 Q^2}{x}\int \frac{dx}{r^2(x)} \ ,
\end{equation}
which can be combined with (\ref{cor}) to obtain the exact form of $A(x)$. Before doing that, it is useful to consider some particular situations. In the far limit, $x\to\infty$, and regardless of the sign of $\epsilon$, this expression boils down to 
 \begin{equation}\label{eq:A(x)-far}
 	A\approx 1-\kappa^2\eta^2-\frac{2M_0}{r}+\frac{\kappa^2Q^2}{r^2} \qquad\text{with}\quad \Omega_{+}=1 \ ,
 \end{equation}
and recovers the  Reissner-Nordstr\"om solution with topological charge expected in GR \cite{Jusufi2016, Prasetyo2017} (this requires setting to zero an integration constant). The structure of horizons, therefore, will be similar to the Reissner-Nordstr\"om case if the topological charge is sufficiently small. 

The other limit of interest corresponds to $x\to 0$. A glance at Eq. (\ref{cor}) indicates that for  $\epsilon<0$ the function $r^2(x)$ has a minimum value given by 
  \begin{equation}\label{minimorm}
	r_{min}^2\equiv \frac{1}{2}\left(|\epsilon|\kappa^2\eta^2+\sqrt{|\epsilon|^2\kappa^4\eta^4+4|\epsilon|\kappa^2Q^2}\right) \ ,
  \end{equation}
which occurs  at $x=0$. If $x$ is extended to negative values, $r^2(x)$ grows again, defining a (symmetric) wormhole structure.   On the other hand, for $\epsilon>0$ the situation is quite different because it is the function $x^2(r)=r^2+\epsilon\kappa^2\eta^2+\epsilon \kappa^2Q^2/r^2$ which attains a minimum of magnitude $x^2_{min}=\epsilon \kappa^2\eta^2+2|Q|\kappa\epsilon^{\frac{1}{2}}$ when $r^4=\epsilon \kappa^2Q^2$. Thus here the wormhole structure seems to arise in the geometry associated to $h_{\mu\nu}$. Moreover, given that the  two-spheres of $h_{\mu\nu}$ grow  without bound in the limits $r\to  \infty$ and $r\to0$, it is evident that the wormhole structure in this case is asymmetric. The situation, however, is more subtle because at $r^4=\epsilon \kappa^2Q^2$ the function $\Omega_+$ in the line element (\ref{metrica}) vanishes when $\epsilon>0$, implying that this hypersurface is null. One can check that at this location the $g_{tt}=-A/\Omega_+$ component diverges, which indicates that the Killing vector $\xi=\partial_t$ has divergent norm there. This anomaly can be cured by considering a different normalization for this vector, such as $\tilde{\xi}=\Omega_+\partial_t$. In this case, at infinity $\tilde{\xi}$ coincides with ${\xi}$ but its norm is finite everywhere. Given that the norm of $\tilde{\xi}$  now vanishes at $r=r_c$ this null hypersurface can certainly be regarded as a Killing horizon. 

From a physical perspective, since the matter fields are coupled to the metric $g_{\mu\nu}$, the relevant thing to consider is the behavior of the metric $g_{\mu\nu}$ as $r\to 0$. In this limit (and for $\epsilon>0$), we have $x\approx  \kappa|Q|\epsilon^{\frac{1}{2}}/r$, which leads to $A(x)\approx \kappa^2 Q^2/3 r^2$. This behavior is essentially the same (up to a constant) as one finds in a standard  Reissner-Nordstr\"om black hole in GR when $r\to0$, where $A(r)_{GR}\approx \kappa^2 Q^2/2 r^2$. Thus, the $\epsilon>0$ case is geodesically incomplete (because radial null geodesics hit the central singularity in finite affine time).

Let us now consider the $x\to 0$ limit in the $\epsilon<0$ case. In this region we find $r^2(x)\approx r_{min}^2+r_{min}^2 x^2/(2r_{min}^2-|\epsilon|\eta^2)$, obtaining the approximated expression 
 \begin{equation}\label{eq:A(x)-near}
 A\approx 1-\kappa^2\eta^2-\frac{2M_0}{x}-\frac{\kappa^2Q^2}{r_{min}^2}-\frac{\kappa^2Q^2 C}{x} \ ,   
 \end{equation}
 where $C$ is an integration constant with relevant physical implications, since it controls the behavior of the metric as $x\to 0$. In particular, if $C>-2M_0/\kappa^2 Q^2$, then $\lim\limits_{x\to 0} A(x)= +\infty$, whereas if   $C<-2M_0/\kappa^2 Q^2$, then $\lim\limits_{x\to 0} A(x)= -\infty$.  In the particular case in which $C=-2M_0/\kappa^2 Q^2$, then the metric is finite at the origin, taking the value $A(0)=1-\kappa^2\eta^2+\frac{\kappa^2Q^2}{r_{min}^2}$. This shift in the apparent topological charge would have an impact in the deficit angle of the geometry and could, in principle, be observable. 
 
 The implications of $C\neq -2M_0/\kappa^2 Q^2$ can be derived from the geodesic equation (\ref{Eq}) considering the approximation (\ref{eq:A(x)-near}). Dividing (\ref{Eq})  by $\Omega_+^2$ and interpreting $V_{eff}=\frac{A}{\Omega_+}\left(\frac{L^2}{r^2(x)}+k \right)$ as an effective potential, it is easy to see that if $C+2M_0/\kappa^2 Q^2<0$ then for particles with $k=1$ or $L^2>0$ approaching $x\to 0$ from the right ($x>0$) the potential barrier diverges, implying that the right-hand side vanishes at some finite $x>0$ for every given value of the energy $E$. Accordingly, those particles will bounce before reaching the wormhole throat, staying in the $x>0$ region safely. However, particles approaching from the left ($x<0$) will see an infinite potential well attracting them towards the throat. The problem comes when they attempt to continue their path into the $x>0$ region, because they find an infinite potential barrier, which prevents them from going through, thus causing an undesired physical situation in which all such geodesics terminate. On physical grounds, therefore, one should restrict the validity of the solution to the region $x>0$.  The opposite situation happens if $C+2M_0/\kappa^2 Q^2>0$, with particles from $x>0$ feeling a growing attraction towards $x\to 0$ to face suddenly an infinite potential wall that prevents their transmission into the $x<0$ region. Thus, the most favorable physical situation is that in which $C=-2M_0/\kappa^2 Q^2$ because then all particles with energy $E^2>(1-\frac{r_{min}^2}{\epsilon})(\frac{L^2}{2r_{min}^2}+\frac{k}{2})$ can move freely from one side of the wormhole to the other. The traversable wormhole case, obviously, corresponds to configurations with $\frac{r_{min}^2}{\epsilon}<1$. For $\frac{r_{min}^2}{\epsilon}>1$ we have an horizon, and for $\frac{r_{min}^2}{\epsilon}=1$ we find an extremal situation. 
 
\begin{figure}[h]
	\centering
	\includegraphics[height=5cm]{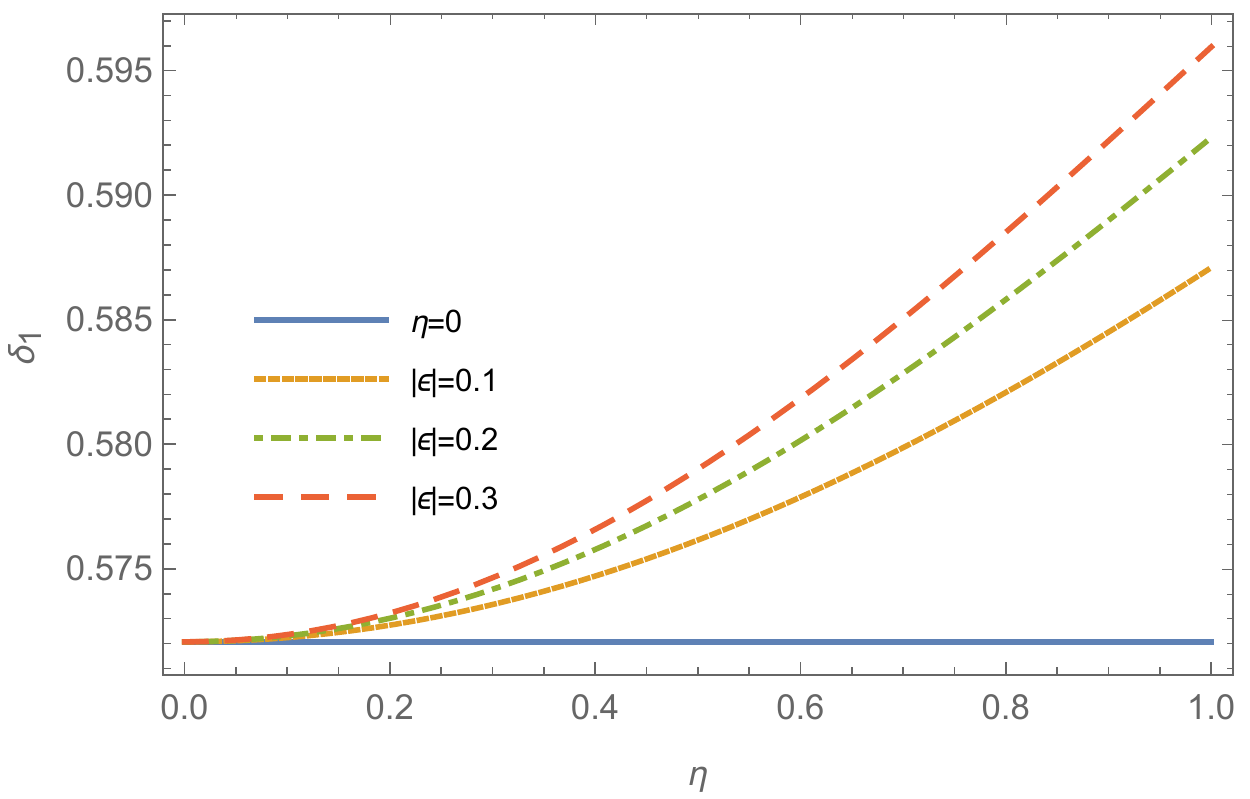}
	\caption{The ``regularization" constant $\delta_1$, for the case $Q=1$, $\kappa^2=1$.} 
	\label{reg}
\end{figure}

From the approximations made so far, the explicit numerical value of the constant $C$ remains completely undetermined. In order to improve this situation, it is useful to look for an exact solution of Eq. (\ref{eq:A(x)}) which may shed light on this parameter. To progress in this direction, one can use the relationship {(\ref{cor})} between the auxiliary coordinate $x$ and the radial coordinate $r$, to show that (\ref{A}) can be written as
\begin{equation}\label{A2}
	A=1-\frac{2M_0r}{\sqrt{r^4-\kappa^2|\epsilon|(\eta^2r^2+Q^2)}}-\frac{\kappa^2Q^2+\kappa^2\eta^2r^2}{3r^2}-\frac{2r}{3\sqrt{r^4-\kappa^2|\epsilon|(\eta^2r^2+Q^2)}}I(r),
\end{equation}
{where} $I(r)=I_1(r)+I_2(r)$, {and}
\begin{equation}\label{int}
	I_1(r)=\int\frac{2\kappa^2Q^2
		}{\sqrt{r^4-\kappa^2|\epsilon|(\eta^2r^2+Q^2)}}dr,\quad	I_2(r)=\int\frac{\kappa^2\eta^2r^2
	}{\sqrt{r^4-\kappa^2|\epsilon|(\eta^2r^2+Q^2)}}dr.
\end{equation}
Taking the proper limits ($Q=0$ or $\eta=0$) we directly retrieve the cases already studied \cite{Nordstrom-olmo,Shaikh2015,lambaga2018gravitational}. To simplify the analysis, it is useful to write $I(r)$ in terms of the minimum radius $r_{min}$, which leads to
\begin{equation}
		I_1(r)=\int\frac{2\kappa^2Q^2
		}{r^2\sqrt{\left(1-\frac{r_{min}^2}{r^2}\right)\left(1+\frac{\kappa^2|\epsilon|Q^2}{r_{min}^2r^2}\right)}}dr \ , \ \quad	I_2(r)=\int\frac{\kappa^2\eta^2
	}{\sqrt{\left(1-\frac{r_{min}^2}{r^2}\right)\left(1+\frac{\kappa^2|\epsilon|Q^2}{r_{min}^2r^2}\right)}}dr.
\end{equation}
{The explicit results for these integrals are}
\begin{eqnarray}
	I_1&=&-\frac{2\kappa^2Q^2}{r}AppellF_1\left[\frac{1}{2},\frac{1}{2},\frac{1}{2},\frac{3}{2},\left(\frac{r_{min}}{r}\right)^2,-\frac{\kappa^2|\epsilon|Q^2}{(rr_{min})^2}\right],\\
	I_2&=&\kappa^2\eta^2rAppellF_1\left[-\frac{1}{2},\frac{1}{2},\frac{1}{2},\frac{1}{2},\left(\frac{r_{min}}{r}\right)^2,-\frac{\kappa^2|\epsilon|Q^2}{(rr_{min})^2}\right].
\end{eqnarray}
The condition to avoid divergences at the throat of the wormholes is that
\begin{equation}\label{re}
	2M_0=-\frac{2}{3}I(r_{min}) \ ,
\end{equation}
which was already observed in  \cite{Shaikh2015}. 
With this condition (\ref{re}), {it} can be shown that  (\ref{A2}) is regular {at} $r=r_{min}$ and tends to 
\begin{equation}\label{A0}
	\lim\limits_{r\to r_{min}}A=A_0=1-\frac{\kappa^2\eta^2}{2}-\frac{\sqrt{\kappa^4|\epsilon|^2\eta^4+4\kappa^2|\epsilon|Q^2}}{2|\epsilon|}=1-\frac{r_{min}^2}{|\epsilon|}.
\end{equation}
The relation (\ref{A0}) means that if $\frac{r_{min}^2}{|\epsilon|}>1$ then the minimal surface will be hidden behind an event horizon, which will be regular. If $\frac{r_{min}^2}{|\epsilon|}<1$,  then the solution describes a traversable wormhole. And when $\frac{r_{min}^2}{|\epsilon|}=1$, we have an extremal black hole. 
By expanding around $r_{min}$, one finds that $I_2$ vanishes, leading to  $2M_0=-\frac{2}{3}I(r_{min})$, which can also be written as
\begin{equation}\label{QM}
M_0=\frac{2\kappa^2Q^2}{3r_{min}}K\left[-\frac{|\epsilon|\kappa^2 Q^2}{r_{min}^4}\right]+\frac{r_{min}\kappa^2\eta^2}{2}
\left(K\left[-\frac{|\epsilon|\kappa^2Q^2}{r_{min}^4}\right]-E\left[-\frac{|\epsilon|\kappa^2Q^2}{r_{min}^4}\right]
\right),
\end{equation}
where $K[x]$  is the complete elliptic integral of the first kind and $E[x]$ is the complete elliptic integral of the second kind. Note that this expression is related to the constant $C$ introduced above via 
\begin{equation}
\label{valC}
C=-\frac{4}{3r_{min}} K\left[-\frac{|\epsilon|\kappa^2Q^2}{r_{min}^4}\right]-\frac{2r_{min}\eta^2}{3Q^2}\left(K\left[-\frac{|\epsilon|\kappa^2Q^2}{r_{min}^4}\right]-E\left[-\frac{|\epsilon|\kappa^2Q^2}{r_{min}^4}\right]
\right).
\end{equation}
From this expression it follows that if we take $Q=0$, the condition (\ref{QM}) implies $M_0=0$. This means that the canonical model only supports regular solutions when the mass is {zero}, as seen in the previous subsection. Now, if we take $\eta=0$, the solution boils down to that  of an electric charge, where $\delta_1\equiv\frac{\kappa^2Q^2}{2M_0r_{min}}\simeq0.572$  \cite{Nordstrom-olmo,Shaikh2015}. When one allows for the simultaneous coexistence of both charges ($\eta\neq0$, $Q\neq0$), the effect of the GM charge translates into a ``regularization" constant $\delta_1$ bigger than that predicted in the pure electric case, as illustrated in  Fig.\ref{reg}. 

\begin{figure}[h]
	\centering
	\includegraphics[height=5.5cm]{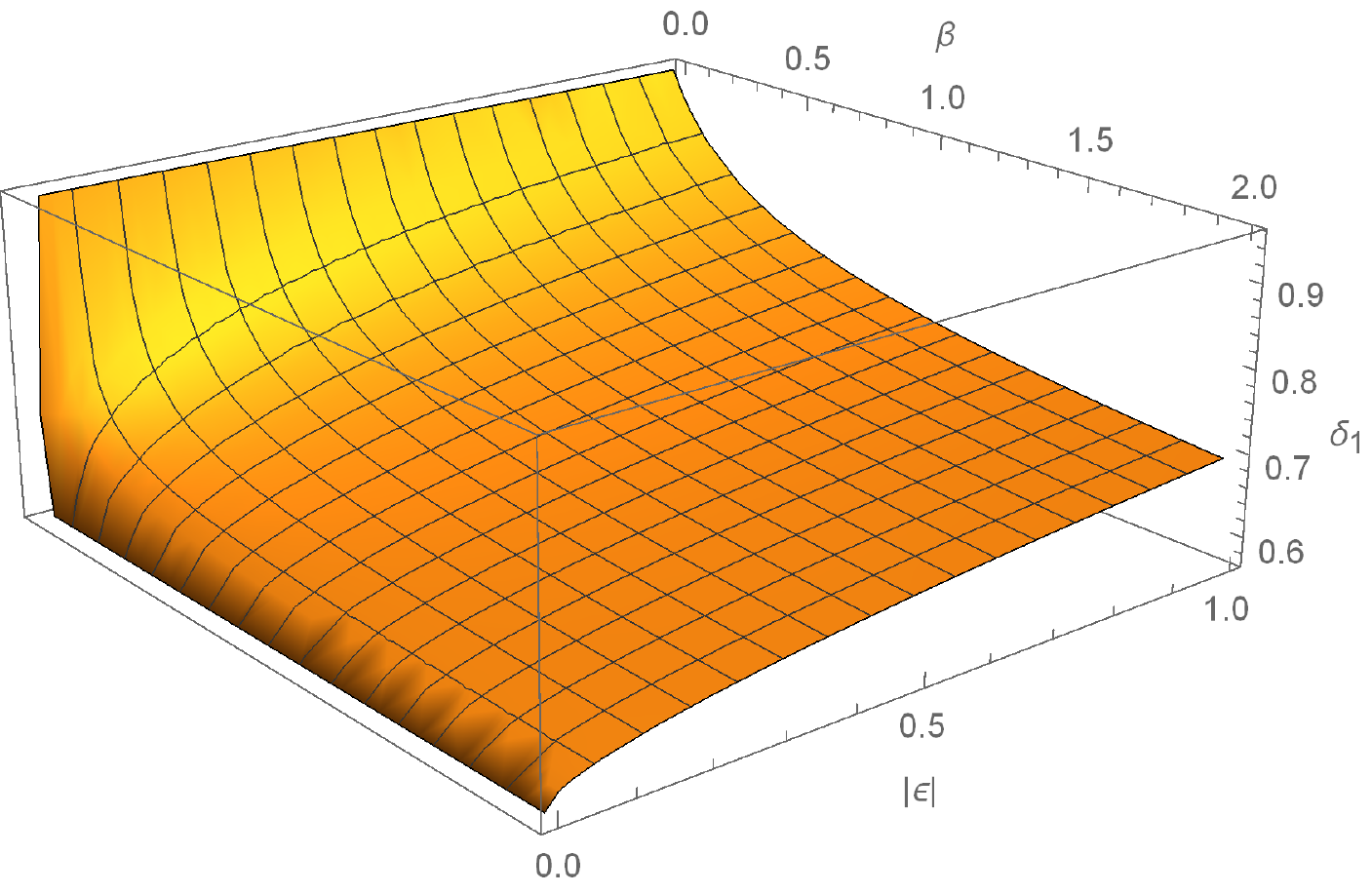}
	\caption{Representation of the function $\delta_1$ in terms of $\eta$ and $\beta$.} 
	\label{regeta}
\end{figure}

\begin{figure}[h]
	\centering
	\includegraphics[height=5cm]{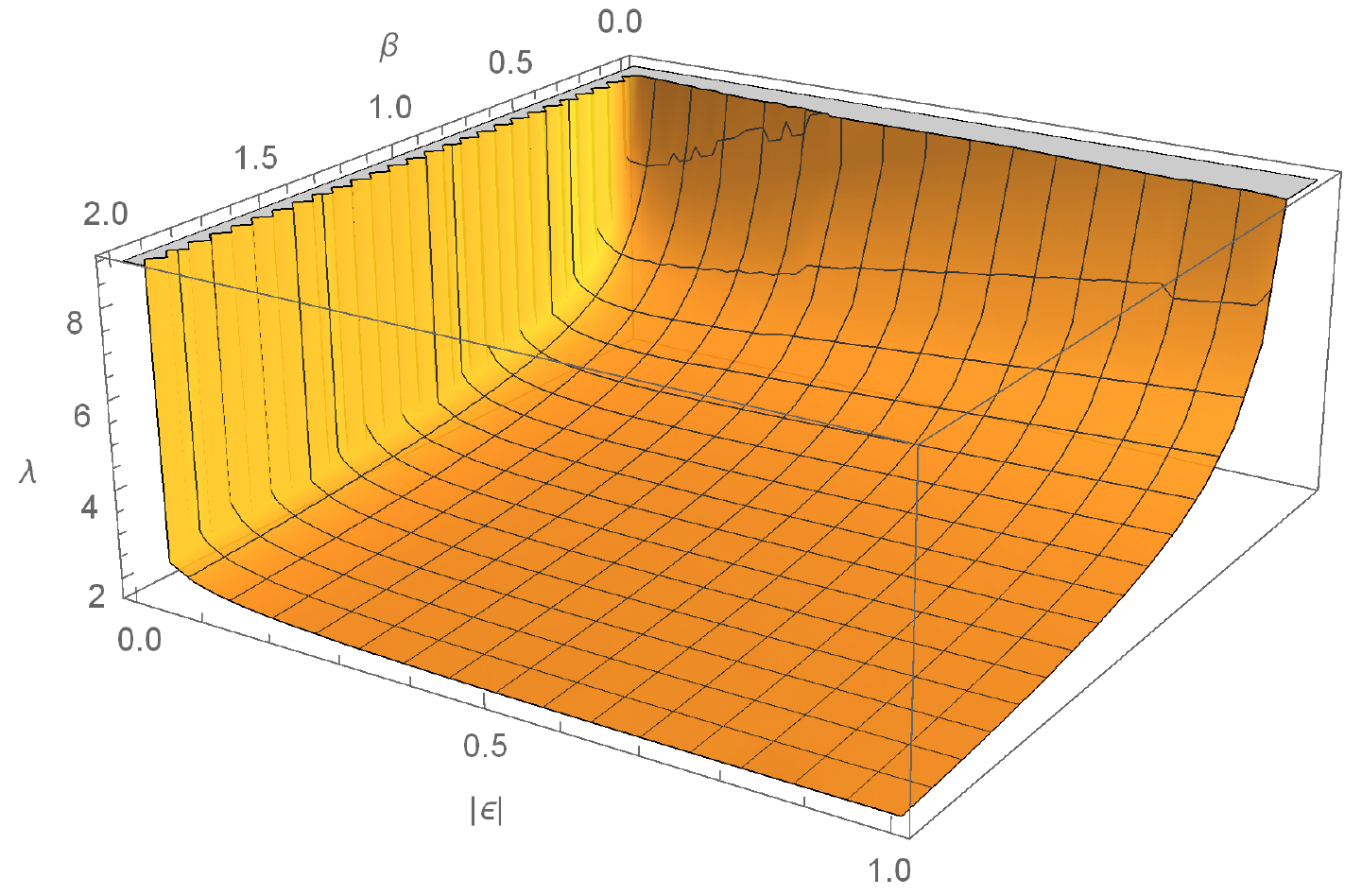}
	\caption{Representation of $\lambda$ as a function of $\beta$ and $\epsilon$.} 
	\label{lambda}
\end{figure}

Let us go back to the original formulation of the problem and restore the dependence of $Q$ on $\eta$.  In this case the constant $\delta_{1}$ is given by 
\begin{equation}
	\delta_1=\left[
	\frac{4}{3} K\left[-\frac{|\epsilon|\kappa^2Q^2}{r_{min}^4}\right]+\frac{2r^2_{min}\eta^2}{3Q^2}\left(K\left[-\frac{|\epsilon|\kappa^2Q^2}{r_{min}^4}\right]-E\left[-\frac{|\epsilon|\kappa^2Q^2}{r_{min}^4}\right]
\right)
	\right]^{-1},
\end{equation}
where $Q^2=\beta \eta^4$, and it is apparent that for a given model, characterized by a  pair $(\epsilon,\beta)$, the value of $\delta_1$ is fixed.  To see how $\delta_1$ depends on $\epsilon$ and $\beta$, in Fig. \ref{regeta} we plot $\delta_{1}(\beta,|\epsilon|)$, where $\lim\limits_{\beta\to\infty}\delta_{1} \simeq 0.57$, $\lim\limits_{\beta\to0}\delta_{1} \simeq 0.95$, $\lim\limits_{|\epsilon|\to0}\delta_{1} \simeq 0.57$,  $\lim\limits_{|\epsilon|\to \infty}\delta_{1} \simeq 0.95$. This allows us to conclude that $\delta_1$ increases when $|\epsilon|$ grows and decreases when  $\beta$ grows.
From the relation (\ref{QM}) we can also {find} how the charge-to-mass ratio ($\eta/M_0$) determines what kind of solution we will have. The relevant expression to use is the following:
\begin{equation}\label{etaM}
	\frac{\eta}{M_0}=\frac{\lambda}{r_{min}^2/|\epsilon|}, \quad\text{where}\quad\lambda=\frac{3}{|\epsilon|\kappa^2}
	\left[\frac{2\beta}{\xi^{3/2}}K\left[-\frac{\kappa^2|\epsilon|\beta}{\xi^2}\right]+\frac{1}{\xi^{1/2}}\left(K\left[-\kappa^2\frac{|\epsilon|\beta}{\xi^2}\right]-E\left[-\frac{\kappa^2|\epsilon|\beta}{\xi^2}\right]\right)
	\right]^{-1},
	%\frac{\left(\frac{|\epsilon|+\sqrt{|\epsilon|^2+4|\epsilon|\beta}}{2}\right)^{3/2}}{K\left[-\frac{4|\epsilon|\beta}{(|\epsilon|+\sqrt{|\epsilon|^2+4|\epsilon|\beta})^2}\right]} \ ,
\end{equation}
which is plotted in Fig. \ref{lambda}, here $\xi=\frac{|\epsilon|\kappa^2+\sqrt{|\epsilon|^2\kappa^4+4|\epsilon|\kappa^2\beta}}{2}$. There we see $\lambda$ {as a function of various parameters} of the model. As one can see, $\lambda$ decreases when the values of $|\epsilon|$ and $\beta$ increase. In the context of the discussion below Eq. (\ref{A0}), we see that  (\ref{etaM}) allows us to conclude that if $\frac{\eta}{M}>\lambda$, we will have a traversable wormhole, but if $\frac{\eta}{M}<\lambda$ then we have a regular black hole. We thus conclude that the further we move away from GR and the canonical NLSM the less GM charge is required to obtain regular solutions. Nonetheless, recall that the regularity of the solution is only possible thanks to the nonlinearity of the NLSM kinetic term.

   \section{Conclusion}\label{conclusao}

	In this work we have considered two different nonlinear $\sigma$-models  minimally coupled to EiBI gravity as a way to explore new gravitational phenomena related with a global monopole symmetry breaking ($SO(3)\to U(1)$).  The first model, which we refer to as canonical model, simply reproduces the analysis of \cite{lambaga2018gravitational} but unveils new features of the solutions besides considering another branch of the allowed parameters of the gravity theory. The second model describes a theory with quadratic kinetic term and leads to an effective configuration closely related to an electric field coupled to a global monopole.  
	
	As a general feature, we found that for the two models considered the asymptotically far solution coincides with that of GR but 
wormhole geometries arise with different peculiarities as one approaches the central region. When $\epsilon<0$ the wormhole is associated to the physical metric, $g_{\mu\nu}$, while for $\epsilon>0$ it is associated to the auxiliary metric $h_{\mu\nu}$. Though none of the models considered managed to yield completely regular space-times, the power law model showed signs of improvement in this direction. In fact, in that case there exists a family of solutions for which the metric is finite everywhere and allows particles to go from one side of the wormhole to the other. This happens when the charge-to-mass ratio satisfies a specific constraint  encoded in the integration constant $C$, which must take the value given by (\ref{valC}). When $C$ is smaller or greater than this value, the geodesics of massive particles can terminate on the wormhole throat, thus leading to a physically undesirable situation. Something similar happens in the canonical model studied when $\epsilon<0$. This provides additional evidence to the fact that wormholes do not necessarily guarantee the completeness of geodesics. 
	
	The regular case somehow generalizes the solution found in \cite{Nordstrom-olmo} to cases with nonzero topological charge. We have shown that the parameter $\delta_1$, which is related to the constant $C$, is bounded below by the purely electric case. This parameter controls the traversability of the wormhole, in the sense that depending on the charge-to-mass ratio of the monopole charge an event horizon may be present or not. In  \cite{Nordstrom-olmo} the wormhole is always traversable even for massive particles, without the pathologies found here regarding geodesic motion. 
	
	An important remark concerning the nature of the solutions is in order. In GR, solutions with a global monopole corresponding to the ansatz considered here have a divergence in $\partial_\mu \phi^a$ at $r=0$, where the space-time exhibits a conical singularity generated by the topological defect \cite{Kibble}. In our modified gravity context, the existence of a wormhole at finite $r$ (if $\epsilon<0$) indicates that those fields must have a topological origin, not being related to any defects but to topological fluxes (even in the case with $\epsilon>0$, for which the wormhole throat lies at $r=0$). The charges associated to the scalar fields, therefore, can be seen as emergent properties of the wormholes, as topological virtues rather than defects, which allows to interpret them as geons in Wheeler's sense \cite{geons}. Further analyses in this direction are currently underway and will be reported elsewhere. 
	
	To conclude, it should be noted that the solutions found here generated by scalar fields (nonlinear sigma models) have astrophysical implications completely different from those found in the case of coupling EiBI to a massless free field \cite{Riccibased}. In that case, compact solutions strongly modify the geometry far from the center, possibly having a substantial astrophysical impact as compared to the predictions of GR. In the case considered here, the geometry is only substantially modified in the innermost regions close to the center, in much the same way as it happens with electric fields. This indicates that the nonlinear dynamics of EiBI gravity can manifest itself in very different ways, depending intimately on the type and properties of the matter fields coupled to it. 
	
%\vspace*{2mm}

\section*{Acknowledgments}
G. J. O. is funded by the Ramon y Cajal contract RYC-2013-13019 (Spain).  This work is supported by the Spanish project FIS2017-84440-C2-1-P (MINECO/FEDER, EU), the project SEJI/2017/042 (Generalitat Valenciana), the project i-LINK1215 (CSIC), and the Severo Ochoa grant SEV-2014-0398 (Spain).  The work by A. Yu. P. has been supported by the CNPq project No. 303783/2015-0.  PJP would like to thank the Brazilian agency CAPES for financial support (PDE/CAPES grant, process 88881.17175/2018-01) and Department of Physics and Astronomy, University of Pennsylvania, for the hospitality.

\vspace*{3mm}

%\renewcommand{\thesection}{}		
%\centerline{\bf APPENDIX. TOPOLOGICALLY CHARGED ELLIS WORMHOLE}
\appendix{\bf Topologically charged Ellis wormhole}

%  \subsubsection{Topologically charged Ellis wormhole}

\vspace*{3mm}

 We saw that the condition $M_0=0$ provides a traversable wormhole, well behaved everywhere and as simple as that of Morris and Thorne \cite{MorrisThorne}, with the difference that we have a solid angle deficit. Such a scenario can be interpreted as a result of the complete evaporation of the black hole, therefore, only the topological charge does not disappear \cite{Pramana}. Next, let us take $M_0=0$ and $\epsilon<0$ to analyze this case in more detail. Thus, the (\ref{Ac})  can be represented as
\begin{equation}\label{wm}
ds^2=-dt^2+\frac{dr^2}{(1-\kappa^2\eta^2)\left(1-\frac{|\epsilon|\kappa^2\eta^2}{r^2}\right)}+r^2(d\theta^2+\sin^2\theta d\phi^2),
\end{equation}
where we have absorbed the factor $(1-\kappa^2\eta^2)$ into a rescaled time coordinate.
This is the metric of a wormhole with a deficit solid angle. Comparing with the Morris-Thorne metric given by
\begin{equation}
ds^2=-e^{\Phi(r)}dt^2+\frac{dr^2}{1-\frac{b(r)}{r}}+r^2(d\theta^2+\sin^2\theta d\phi^2),
\end{equation}
we identified that the redshift function is zero, $\Phi(r)=0$, and the shape function is given by
\begin{equation}
b(r)=r\kappa^2\eta^2+(1-\kappa^2\eta^2)\frac{|\epsilon|\kappa^2\eta^2}{r}.
\end{equation}
We note that $\lim\limits_{r\to\infty}\frac{b(r)}{r}=\kappa^2\eta^2$. Thus, the solution (\ref{wm})  is not asymptotically flat \cite{phantomwormholesolution2013}, which is natural due to the GM charge. We can build the embedding diagram. For this, it is enough to consider the  time constant and the equatorial plane, that is, $\theta=\frac{\pi}{2}$. 
\begin{equation}\label{me1}
ds^2=\frac{dr^2}{(1-\kappa^2\eta^2)\left(1-\frac{|\epsilon|\kappa^2\eta^2}{r^2}\right)}+r^2d\phi^2.
\end{equation}
Now we want to construct a surface in three-dimensional Euclidean space with the same characteristics as the metric above, for this we will consider the three-dimensional metric in cylindrical coordinates ($z,r,\phi$): $ds_{E}^2=dz^2+dr^2+r^2d\phi^2$, where we have
\begin{equation}\label{me2}
ds_{E}^2=\left[1+\left(\frac{dz}{dr}\right)^2\right]dr^2+r^2d\phi^2.
\end{equation}
Identifying the metrics (\ref{me1}) and (\ref{me2}) , we have
\begin{equation}\label{eqI}
\frac{1}{r_{min}^2}\left(\frac{dz}{dy}\right)^2=\frac{1}{(1-\kappa^2\eta^2)\left(1-\frac{1}{y^2}\right)}-1,
\end{equation}
where $r_{min}=\sqrt{|\epsilon|}\kappa\eta$  is the radius of the throat and $y=\frac{r}{r_{min}}$.  Evaluating numerically the equation (\ref{eqI}), we get the function $z(y)/r_{min}$ given by{ Fig. \ref{t} and the immersion diagram depicted at Fig. \ref{WH}.
\begin{figure}[h]
	\centering
	\includegraphics[height=5cm]{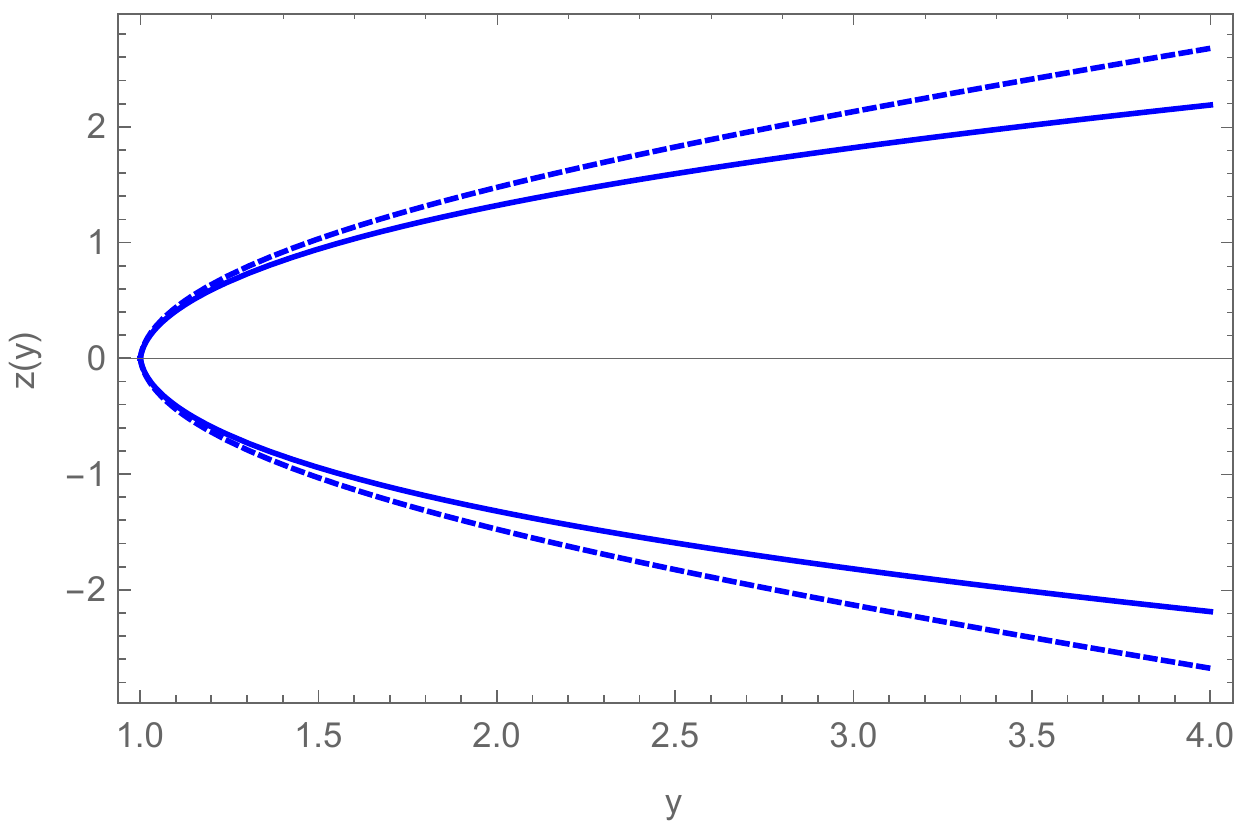}
	\caption{$z(y)/r_{min}$. The full line corresponds to $\kappa\eta=0.2$ and the dotted line, $\kappa\eta=0.4$} 
	\label{t}
\end{figure}

\begin{figure}[h]
	\centering
	\includegraphics[height=5cm]{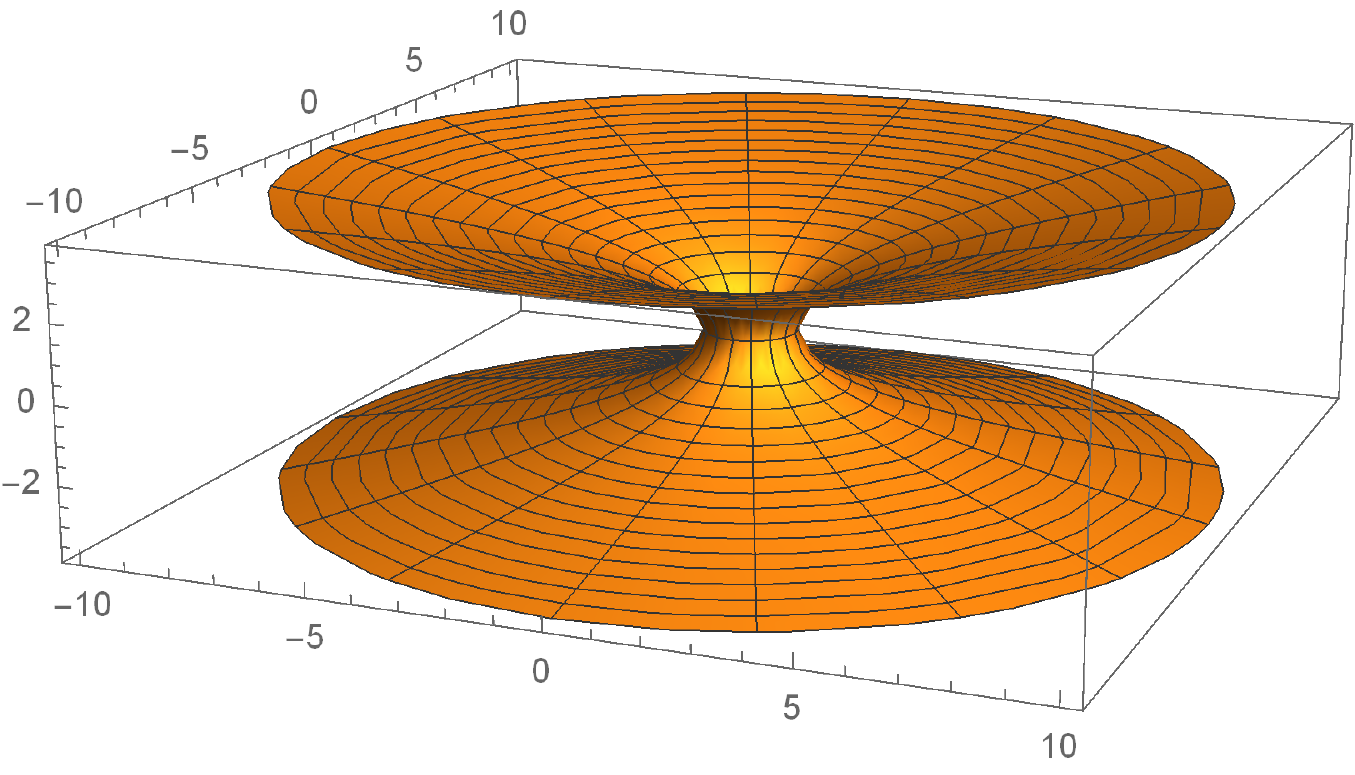}
	\caption{Embedding diagram of a section $\theta=\frac{\pi}{2}$ and $t=\text{constant}$. We are take  and $\kappa\eta=0.2$. The blue side corresponds to $x>0$ and the yellow side corresponds to $x<0$. } 
	\label{WH}
\end{figure}

Notice that the shape of the wormhole is conical, due to the GM charge. As far as we know, this is the simplest wormhole solution found in EiBI gravity. In \cite{Kimet2018}, Jusufi also obtained a static and asymptotically conical wormhole, but in the GR context. For this, he considered the minimal coupling of the GM tensor ($T^{\mu}_{\phantom{\mu}\nu}=diag\left(-\eta^2/r^2,-\eta^2/r^2, 0,0 \right)$ to the gravity} plus an anisotropic fluid. The fluid obeys the state equation of the form $\mathcal{P}_r(r)=\omega\rho(r)$ com $\omega<1$, corresponding to the phantom energy as it must be for the wormhole-like solutions. 
 As we can see, the model describes a triplet of scalar fields whose topological charges depend on the the side of the wormhole where the observer is positioned. At $x>0$, the topological charge is positive while at $x<0$ it is negative. In other words, observers at  $x>0$ ($x<0$) face a some kind of a global monopole (anti-monopole) \cite{Rajaraman}.


\begin{thebibliography}{50}
	
	
	
	
 \bibitem{Kibble}T. W. B. Kibble, J. Phys. A {\bf 9}, 1387 (1976).
 
 \bibitem{GM}M. Barriola and A. Vilenkin, Phys. Rev. Lett. {\bf 63}, 341 (1989).
 
  \bibitem{Pramana} N. Dadhich, K. Narayan, and U. A. Yajnik, Pramana J. Phys.
 	50, 307 (1998).%Schwarzschiid black hole with global monopole charge 
 	
  \bibitem{Man2011} H. Cheng and J. Man, Class. Quant. Grav. {\bf28},
 		015001 (2011).
 
 \bibitem{cosmology}  Supernova Search Team Collaboration, A. G. Riess et al., Astron. J. {\bf 116}, 1009 (1998);
  Supernova Cosmology Project Collaboration, S. Perlmutter et al., Astrophys. J. {\bf 517}, 565 (1999); Boomerang Collaboration, C. B. Netterfield et al., Astrophys. J. {\bf 571}, 604 (2002);  D. La, P. J. Steinhardt, Phys. Rev. Lett. {\bf 62}, 376 (1989).
 
 
 \bibitem{Vollick:2003qp}
  D.~N.~Vollick,
  %``Palatini approach to Born-Infeld-Einstein theory and a geometric description of electrodynamics,''
  Phys.\ Rev.\ D {\bf 69}, 064030 (2004)
  %doi:10.1103/PhysRevD.69.064030
  [gr-qc/0309101].
  
  \bibitem{banados2010eddington}  M. Ba\~nados and P. G. Ferreira, Phys. Rev. Lett. {\bf 105}, 011101 (2010).
  
  

  \bibitem{eddington1920mathematical} A. S. Eddington, The Mathematical Theory of Relativity,
  (Cambridge University Press, Cambridge, England, 1924).
  
  \bibitem{born1934foundations} M. Born and L. Infeld, % Foundations of the new field theory
  Proc. R. Soc. A {\bf 144}, 425 (1934).
  
 
   \bibitem {BeltranJimenez:2017doy} J. B. Jimenez, L. Heisenberg, G. J. Olmo, and D. Rubiera-Garcia, %Born-Infeld inspired modifications of gravity,
    Phys. Rep. {\bf 727}, 1 (2018).
    
    \bibitem{Olmo:2015dba} 
  G.~J.~Olmo, D.~Rubiera-Garcia and A.~Sanchez-Puente,
  %``Classical resolution of black hole singularities via wormholes,''
  Eur.\ Phys.\ J.\ C {\bf 76}, no. 3, 143 (2016). 
  
  \bibitem{Bazeia:2015uia} 
  D.~Bazeia, L.~Losano, G.~J.~Olmo, D.~Rubiera-Garcia and A.~Sanchez-Puente,
  %``Classical resolution of black hole singularities in arbitrary dimension,''
  Phys.\ Rev.\ D {\bf 92}, no. 4, 044018 (2015).
  
  \bibitem{Olmo:2015bya} 
  G.~J.~Olmo, D.~Rubiera-Garcia and A.~Sanchez-Puente,
  %``Geodesic completeness in a wormhole spacetime with horizons,''
  Phys.\ Rev.\ D {\bf 92}, no. 4, 044047 (2015)
  
  \bibitem{Bambi:2015zch} 
  C.~Bambi, A.~Cardenas-Avendano, G.~J.~Olmo and D.~Rubiera-Garcia,
  %``Wormholes and nonsingular spacetimes in Palatini $f(R)$ gravity,''
  Phys.\ Rev.\ D {\bf 93}, no. 6, 064016 (2016).
    
    
    \bibitem{barros1997global} A. Barros and C. Romero, Phys. Rev. D {\bf 56}, 6688 (1997).
    
    \bibitem{carames2011gravitational} T. R. P. Caram\^es, E. R. Bezerra de Mello, M. E. X. Guimar\~aes, Int. J. Mod. Phys. D: Conf. Ser. {\bf 3}, 446 (2011).
    
     \bibitem{carames2012motion} T. R. P. Caram\^es, E. R. Bezerra de Mello, M. E. X. Guimar\~aes, Mod. Phys. Lett. A {\bf 27}, 1250177 (2011).
     
     \bibitem{morais2012gravitational} J. P. Morais da Gra\c{c}a and V. B. Bezerra, Mod. Phys. Lett. A {\bf 27}, 1250178 (2012).
     
     \bibitem{man2013thermodynamic} Jingyun Man and Hongbo Cheng, Phys. Rev. D {\bf 87}, 044002 (2013).
     
     \bibitem{man2015analytical} Jingyun Man and Hongbo Cheng, Phys. Rev. D {\bf 92}, 024004 (2015).
     
     \bibitem{carames2017f} T. R. P. Caram\^es, J. C. Fabris, E.R. Bezerra de Mello, H. Belich, Eur. Phys. J. C {\bf 77}, 496 (2017).
     
    \bibitem{Palatinef(R)mono} J. R. Nascimento, G. J. Olmo, A. Yu. Petrov, P. J. Porfirio, A. R. Soares,  Phys.\ Rev.\ D {\bf 99}, 064053 (2019).
    
    \bibitem{lambaga2018gravitational} R. D. Lambaga and H. S. Ramadhan, Eur. Phys. J. C {\bf 78}, 436 (2018).
    % Gravitational field of global monopole within the Eddington-inspired Born-Infeld theory of gravity
    
     \bibitem{Hongwei}Hongwei Tan, Jinbo Yang, Jingyi Zhang and Tangmei He, Chin. Phys. B {\bf 27}, 3 (2018). 
    
     \bibitem{olmo2011} G. J. Olmo, Int. J. Mod. Phys. D {\bf 20}, 413 (2011).
     % ``Palatini Approach to Modified Gravity: f(R) Theories and Beyond,''[arXiv:1101.3864 [gr-qc]].
     
     
     \bibitem{BeltranJimenez:2019acz} 
  J.~Beltr\'{a}n Jim\'{e}nez and A.~Delhom,
  %``Ghosts in metric-affine higher order curvature gravity,''
  Eur.\ Phys.\ J.\ C {\bf 79}, 656 (2019).
%  doi:10.1140/epjc/s10052-019-7149-x
%  [arXiv:1901.08988 [gr-qc]]. 
     
     \bibitem{Afonso:2017bxr} 
  V.~I.~Afonso, C.~Bejarano, J.~Beltran Jimenez, G.~J.~Olmo and E.~Orazi,
  %``The trivial role of torsion in projective invariant theories of gravity with non-minimally coupled matter fields,''
  Class.\ Quant.\ Grav.\  {\bf 34}, 235003 (2017).
%  doi:10.1088/1361-6382/aa9151
%  [arXiv:1705.03806 [gr-qc]].
     
     
      \bibitem{Babichev2006} E. Babichev, Phys. Rev. D {\bf74}, 085004 (2006). 
      % Global topological k-defects [hep-th/0608071]
      
      \bibitem{Jin2007} X. H. Jin, X. Z. Li and D. J. Liu, Class. Quant. Grav. {\bf24}, 2773 (2007). 
      %  Gravitating global k-monopole [arXiv:0704.1685 [gr-qc]]
   
    
    \bibitem{Prasetyo2017} I. Prasetyo and H. S. Ramadhan, Gen. Rel. Grav. {\bf49}, 115 (2017).
    
    \bibitem{Gell} M. Gell-Mann and B. Zwiebach, Phys. Lett.  B {\bf141}, 333 (1984).
    % Spacetime compactification induced by scalars 
    
      \bibitem{prep} \textit{In preparation.}
      
       \bibitem{MorrisThorne}M. S. Morris and K. S. Thorne, Am. J. Phys. {\bf 56}, 395 (1988).
       % Wormhole in spacetime and their use for interstellar travel: A tool for teaching general relativity
       
       
       \bibitem{phantomwormholesolution2013} F. S. N. Lobo, F. Parsaei, and N. Riazi, Phys. Rev. D {\bf87}, 084030 (2013).
       % New asymptotically flat phantom wormhole solutions
 
     \bibitem{Rajaraman} R. Rajaraman, Solitons and Instantons: An Introduction to Solitons and Instantons in Quantum Field Theory (North-Holland, 1982). 
    
           
       
       \bibitem{Kimet2018} K. Jusufi, Phys. Rev. D {\bf98}, 044016 (2018).
       % Conical Morris-Thorne wormholes with a global monopole charge-Kimet Jusufi
       
        \bibitem{Nordstrom-olmo} G. J. Olmo and D. Rubiera-Garcia, Phys. Rev. D {\bf86},  044014 (2012);
        % Reissner-Nordstro�m black holes in extended Palatini theories       
%        \bibitem{Olmo:2013gqa} 
  G.~J.~Olmo, D.~Rubiera-Garcia and H.~Sanchis-Alepuz,
  %``Geonic black holes and remnants in Eddington-inspired Born-Infeld gravity,''
  Eur.\ Phys.\ J.\ C {\bf 74}, 2804 (2014)
        
         \bibitem{Jusufi2016}K. Jusufi, Astrophys. Space Sci. {\bf 361}, 24 (2016). 
         % Gravitational lensing by Reissner-Nordstr�m black holes with topological defects
         
         
          \bibitem{Shaikh2015} R. Shaikh, Phys. Rev. D {\bf92}, 024015 (2015). 
          % Lorentzian wormholes in Eddington-inspired Born-Infeld gravity

\bibitem{geons} J.~A.~Wheeler,
  %``Geons,''
  Phys.\ Rev.\  {\bf 97}, 511 (1955);
  G.~J.~Olmo and D.~Rubiera-Garcia,
  %``Geons in Palatini Theories of Gravity,''
  Fundam.\ Theor.\ Phys.\  {\bf 189}, 161 (2017);
  C.~A.~R.~Herdeiro, A.~M.~Pombo and E.~Radu,
  %``Asymptotically flat scalar, Dirac and Proca stars: discrete vs. continuous families of solutions,''
  Phys.\ Lett.\ B {\bf 773}, 654 (2017).
  %doi:10.1016/j.physletb.2017.09.036
 % [arXiv:1708.05674 [gr-qc]].

   
   \bibitem{Riccibased}       
    V.~I.~Afonso, G.~J.~Olmo, E.~Orazi and D.~Rubiera-Garcia,
  %``New scalar compact objects in Ricci-based gravity theories,''
  JCAP {\bf 1912}, 044 (2019).
%  doi:10.1088/1475-7516/2019/12/044
 % [arXiv:1906.04623 [hep-th]].          
          
    %  \bibitem{ShaikhNCC} R. Shaikh, Phys. Rev. D {\bf98}, 064033 (2018).
     
    % \bibitem{Sim-Visser-2019-Bl-bou} Simpson. A and Visser. M, %Black-bounce to traversable wormhole. JCAP \textbf{02} (2019) 042
     
  
   %\bibitem{DBI-2009} D. J. Liu, Y. L. Zhang and X. Z. Li, Eur. Phys. J. C {\bf60} 495 (2009) % A Self-gravitating Dirac-Born-Infeld Global Monopole, [arXiv:0902.1051 [hep-th]].
  
   %\bibitem{DBI-Ddimensao-2018} H. S. Ramadhan, I. Prasetyo, A. M. Kusuma, Gen. Relativ. Gravit. {\bf50} 96 (2018). % Higher-dimensional black holes with Dirac�Born�Infeld (DBI) global defects [arXiv:1807.03944 [gr-qc]].
  
\end{thebibliography}
\end{document}